\documentclass[12pt,a4paper]{article}
\usepackage{amssymb}
\usepackage{amsfonts}
\usepackage{amsmath}

\input{tcilatex}

\begin{document}

\title{A Surprising Property of Multidimensional Hamiltonian Systems;
Application to Semiclassical Quantization of Phase Space }
\author{Maurice de Gosson \\
Blekinge Institute of Technology\\
371 79 Karlskrona, Sweden\\
e-mail: mdg@bth.se}
\maketitle

\section*{Introduction}

In the mid 1980's the mathematician Gromov \cite{Gromov} discovered a very
unexpected property of canonical transformations (and hence of Hamiltonian
motion). That property, often dubbed the \textquotedblleft principle of the
symplectic camel\textquotedblright , seems at first sight to be in conflict
with the common conception of Liouville's theorem; it actually only shows
that canonical transformations have a far more \textquotedblleft
rigid\textquotedblright\ behavior than usual volume-preserving mappings.
Here is one description of Gromov's result (another will be given a moment).
Let us cut a circular hole with radius $r$ in any of the \textit{conjugate}
coordinate planes $q_{j},p_{j}$ in phase-space $\mathbb{R}^{2N}$, and
consider a phase-space ball with radius $R$ larger than $r$. Clearly, we
cannot push the ball through the hole: it is too big. However, we can always 
\textit{squeeze} it through the hole by deforming it using volume-preserving
transformations: conservation of volume does not imply conservation of
shape, and we will be able to turn the ball into, for instance, a long
ellipsoid with smallest half-axis inferior to $r$ and pass it through the
hole. Now, Hamiltonian flows are volume-preserving because of Liouville's
theorem, and we could therefore hope to use such a flow to perform the
squeezing of the ball. But Gromov proved that this is \textit{impossible}
because such a squeezing can \textit{never} be done if one limits oneself to
canonical transformations (and hence, in particular, to Hamiltonian flows):
the proverbial camel will never pass through the eye of the needle if it is
symplectic! Here is an alternative statement of the principle of the
symplectic camel; it has a strong quantum mechanical flavor --although
everything is expressed in classical terms. Consider again a phase-space
ball with radius $R$. Its orthogonal projection on any of the phase-space
planes $q_{i},p_{j}$, $q_{i},q_{j}$, or $p_{i},p_{j}$ is obviously a circle
with area $\pi R^{2}$. Suppose now we let a Hamiltonian flow act on the
ball. It will start distorting; after some time it will perhaps occupy a
very large region of phase-space. Some part of this \textquotedblleft
blob\textquotedblright\ will thus have to become very thin, and one can
hence expect that the areas of at least some of the orthogonal projections
on the coordinate planes will shrink and become very small. However Gromov's
theorem implies that the areas of the projections of the distorted ball on
the \textit{conjugate }planes $q_{j},p_{j}$ will never decrease below their
original value $\pi R^{2}$! This is of course strongly reminiscent of
Heisenberg's uncertainty principle\footnote{%
Especially since the property ceases to hold if $q_{j},p_{j}$ is replaced by
any pair of nonconjugate coordinates.}: suppose that we choose $R=\sqrt{%
\hslash }$; then the area of the projection of the deformed ball on the
conjugate planes will always remain $\geq \frac{1}{2}\hslash $ as time
elapses (see \cite{select} for an explicit derivation of a \textquotedblleft
classical\textquotedblright\ equivalent form Heisenberg inequalities).

Gromov had stumbled onto something big: his theorem has led to a thriving
development of a new field of mathematics, \textit{symplectic topology},
which is related to the theory of periodic orbits for Hamiltonian systems
(see \cite{HZ} and the references therein). The purpose of this Letter is to
show that his principle of the symplectic camel also can be used in physics
to cast some new light on semiclassical quantization, in particular \textit{%
EBK} quantization. This is because the principle of the symplectic camel
allows the definition of a new quantity, the \textit{symplectic area}, which
coincides with the usual notion of area in the two-dimensional case and can
be viewed as a generalization of the notion of \textit{action}. We will show
that symplectic area is a better candidate than volume for the quantization
of phase space; because it allows to recover the usual \textit{EBK} quantum
levels for integrable systems by merely assuming that phase-space is
subdivided in \textquotedblleft quantum blobs\textquotedblright\ with
symplectic area $(n+\frac{1}{2})h$.

\section{Symplectic Area}

Let $\Omega $ be a subset of $\mathbb{R}^{2N}$ ( in our applications it will
be the interior of an energy shell $H(q,p)=E$). We call \textit{symplectic
radius} of $\Omega $, and denote by $R_{\Omega }$, the supremum of all $%
R\geq 0$ such that the phase-space ball 
\begin{equation*}
B(R):|q-q_{0}|^{2}+|p-p_{0}|^{2}\leq R^{2}
\end{equation*}%
can be sent inside $\Omega $ using arbitrary canonical transformations (not
just those arising from Hamiltonians). By definition, the \textit{symplectic
area\footnote{%
Some authors call it "symplectic capacity" but we find this denomination
slightly misleading in our context.} of }$\Omega $ is the number $\func{SA}%
(\Omega )=\pi R_{\Omega }^{2}$. Notice that it can happen that $\func{SA}%
(\Omega )=0$ (no ball can be sent inside $\Omega $) or that $\func{SA}%
(\Omega )=+\infty $ (every ball, no matter its radius, can be sent inside $%
\Omega $). Symplectic area coincides with the usual area in two-dimensional
phase-space $\mathbb{R}^{2}$: the radius of the largest disk that can be
sent inside a surface with area $\pi R^{2}$ using canonical transformations
is precisely $R$ (because canonical transformations are just the area
preserving mappings when $N=1$, as already remarked before), hence the
symplectic area of this region is just its ordinary area $\pi R^{2}$.
However, in higher dimensions symplectic area is not generally directly
linked to volume (see however (\ref{cavol}) below). For instance, the
symplectic area of a ball $B(R)$ in $\mathbb{R}^{2N}$ is obviously 
\begin{equation}
\func{SA}(B(R))=\pi R^{2}  \label{sabr}
\end{equation}%
while its volume is%
\begin{equation}
\func{Vol}B(R)=\frac{\pi ^{N}R^{2N}}{N!}=\frac{1}{N!}(\func{SA}(B(R))^{N}%
\text{.}  \label{cavol}
\end{equation}

An essential observation is that symplectic area is a \textit{symplectic
invariant}: it is conserved by canonical transformations, that is 
\begin{equation}
\varphi \text{ \textit{canonical} }\Longrightarrow \func{SA}(\varphi (\Omega
))=\func{SA}(\Omega )\text{.}  \label{sympre}
\end{equation}%
One can thus say that while volume is the natural invariant for
volume-preserving mappings, symplectic area is the natural invariant for
canonical transformations.

A noticeable feature of the symplectic area of a ball with given radius is
that it is independent of the dimension of the ambient phase-space (as
opposed to its volume). As the number of degrees of freedom increases, the
volume of a ball $B(R)$ decreases towards zero, while its symplectic area
remains equal to $\pi R^{2}$. Also, sets with infinite volume can have
finite symplectic areas. Consider for instance the phase-space cylinder $%
Z_{j}(R)$ with radius $R$ and based on the conjugate coordinate plane $%
q_{j},p_{j}$: a point $(q,p)$ is inside (or on) $Z_{j}(R)$ if and only if
its $j$-th coordinates $q_{j}$ and $p_{j}$ satisfy $q_{j}^{2}+p_{j}^{2}\leq
R^{2}$. The cylinder $Z_{j}(R)$ has infinite volume if $N>1$ (if $N=1$ it is
just a circle in phase-plane), but its symplectic area is $\pi R^{2}$. This
readily follows from Gromov's theorem: suppose we could deform, using
canonical transformations, a ball with radius $R^{\prime }>R$ so that it
becomes a volume $\Omega $ fitting inside $Z_{j}(R)$. We could then also let 
$\Omega $ "fall through the hole\textquotedblright "\ $q_{j}^{2}+p_{j}^{2}%
\leq R^{2}$ in the plane $q_{j},p_{j}$; since translations are canonical
transformations, and the compose of two canonical transformations still is
canonical, we would thus have violated the principle of the symplectic
camel. More generally, using the fact that symplectic area is an increasing
function of size, we see that any subset $\Omega $ of phase-space containing
a ball and which is itself contained in a cylinder $Z_{j}$ with same radius
as the ball will have symplectic area $\pi R^{2}$:%
\begin{equation}
B(R)\subset \Omega \subset Z_{j}(R)\Longrightarrow \func{SA}(\Omega )=\pi
R^{2}\text{.}  \label{inclus}
\end{equation}

We said in the Introduction that symplectic area generalizes the notion of
action. Here is why. Suppose that $\Omega =\Omega (E)$ is bounded by an
energy shell $\Sigma (E):H(q,p)=E$ for some smooth Hamiltonian $H$\footnote{%
In fact every hypersurface $\Sigma $ in phase-space can be viewed as an
energy shell for some Hamiltonian: just choose $H$ equal to $E$ near $\Sigma 
$.}. One shows (\cite{EH,HZ}) that when $\Omega (E)$ is both compact and
convex, then its symplectic area is equal to the lower limit of all the
action integral calculated along periodic Hamiltonian orbits on the energy
shell $\Sigma (E)$:%
\begin{equation}
\func{SA}(\Omega (E))=\inf_{\gamma }\doint\nolimits_{\gamma }pdq\text{ \ \ (}%
\gamma \text{ p.o. on }\Sigma (E)\text{)}  \label{action}
\end{equation}%
with $pdq\equiv p_{1}dq_{1}+\cdot \cdot \cdot +p_{N}dq_{N}$. Moreover, there
exists a minimal periodic orbit $\gamma _{\min }$ for which equality
effectively occurs:%
\begin{equation}
\func{SA}(\Omega (E))=\doint\nolimits_{\gamma _{\min }}pdq\text{.}
\label{min}
\end{equation}%
This property (the proof of which is far from being trivial; see \cite{EH,HZ}%
) does not hold in general if one does not assume $\Omega (E)$ is compact
and connected. Consider for instance a long \textquotedblleft Bordeaux
bottle\textquotedblright\ fitting exactly inside a cylinder $Z_{j}(R)$, and
whose neck has a smaller radius $r<R$. The capacity of the bottle is $\pi
R^{2}$ in view of (\ref{inclus}), but the action of a closed orbit
encircling its neck is $\pi r^{2}<\pi R^{2}$, contradicting formula (\ref%
{min}).

\section{\textquotedblleft Quantum Blobs\textquotedblright\ vs. Quantum Cells%
}

The use of $h^{N}$ as the volume of a \textquotedblleft quantum
cell\textquotedblright\ in phase-space in statistical quantum mechanics is
justified by inference from a few special cases. It turns out that the
consideration of the same particular cases justifies the following
definition:\medskip

\noindent \textbf{Definition}. \textit{A }quantum blob\textit{\ is any
subset of phase space with symplectic area }$(n+\frac{1}{2})h$\textit{,
where }$n$\textit{\ is an integer\ }$\geq 0$.\medskip

\noindent \textbf{Remark 1}. Since symplectic area is a symplectic
invariant, a quantum blob will remain a quantum blob in any system of
canonical coordinates.\medskip

\noindent \textbf{Remark 2}. Notice that dimension does not matter in this
definition: the symplectic area of a quantum blob is independent of which $%
\mathbb{R}^{2N}$ we choose to be a host, as it always is $(n+\frac{1}{2})h$.
Also, as opposed to a quantum cell, a quantum blob can have infinite volume.
For instance, any ball $B(\sqrt{(2n+1)\hbar })$ is a quantum blob, and so
are the cylinders $Z_{j}(\sqrt{(2n+1)\hbar })$ for $j=1,...,N$; the latter
have infinite volume as soon as $N>1$.\medskip

We now make the following \textquotedblleft Quantum Blob
Ansatz\textquotedblright\ (\textit{QBA}): \medskip

\noindent \textbf{Ansatz}: \textit{The only admissible semiclassical motions
are those who take place on the boundary of a quantum blob.\ }\medskip

We are going to see that this Ansatz leads, in spite of its simplicity, to
the correct semiclassical quantum levels for all integrable systems. We
begin by showing, as a first application, that it leads to the correct
quantum features of an ensemble of $N$ linear harmonic oscillators with
collective Hamiltonian%
\begin{equation}
H=\frac{1}{2m}(|p|^{2}+m^{2}\omega ^{2}|q|^{2})=\sum_{j=1}^{N}\frac{1}{2m}%
(p_{j}^{2}+m^{2}\omega ^{2}q_{j}^{2})\text{.}  \label{hiso}
\end{equation}%
The energy shell $\Sigma (E):H(q,p)=E$ is the boundary of the ellipsoid%
\begin{equation}
\frac{1}{2m}(|p|^{2}+m^{2}\omega ^{2}|q|^{2})=E  \label{biso}
\end{equation}%
with interior $\Omega (E)$. Performing the symplectic change of variables%
\begin{equation*}
(q_{j},p_{j})\longmapsto ((m\omega )^{-1/2}q_{j},(m\omega )^{1/2}p_{j})\text{
\ \ }(j=1,...,N)
\end{equation*}%
in (\ref{biso}) the ellipsoid $\Sigma (E)$ becomes the ball 
\begin{equation*}
\frac{\omega }{2}(|p|^{2}+|q|^{2})=E
\end{equation*}%
with radius $\sqrt{2E/\omega }$. Since canonical transformations do not
affect symplectic areas (see (\ref{sympre})) we have%
\begin{equation*}
\func{SA}(\Omega (E))=\limfunc{Cap}B(\sqrt{2E/\omega })=\frac{2\pi E}{\omega 
}
\end{equation*}%
and the requirement that $\Omega (E)$ should be a quantum blob is thus
equivalent to%
\begin{equation*}
\frac{2\pi E}{\omega }=(n+\tfrac{1}{2})h\text{.}
\end{equation*}%
It follows that the energy can only take the values%
\begin{equation*}
E=(n+\tfrac{1}{2})\omega \frac{h}{2\pi }=(n+\tfrac{1}{2})\hbar \omega
\end{equation*}%
and we have thus recovered the energy levels predicted by quantum mechanics.
The volume of the ball $B(\sqrt{2E/\omega })$ being related to its
symplectic area by formula (\ref{cavol}) we have%
\begin{equation*}
\limfunc{Vol}B(\sqrt{2E/\omega })=\frac{1}{N!}\left( \func{SA}(B(\sqrt{%
2E/\omega })\right) ^{N}=\frac{1}{N!}\left( \frac{E}{\hbar \omega }\right)
^{N}
\end{equation*}%
and we hence recover the density of states predicted by statistical
mechanics: 
\begin{equation*}
g(E)=\frac{\partial \func{Vol}(E)}{\partial E}=\left( \frac{1}{\hbar \omega }%
\right) ^{N}\frac{E^{N-1}}{(N-1)!}
\end{equation*}%
where $\func{Vol}(E)$ is the volume of $B(\sqrt{2E/\omega })$.

\section{Quantum Blobs and EBK Quantization}

Let us now broaden the discussion to the more general case of arbitrary $N$%
-dimensional integrable systems; the Hamiltonian is $H=H(q,p)$. We claim
that the \textquotedblleft quantum blob Ansatz\textquotedblright\ leads to
the semiclassical energy levels predicted by semiclassical mechanics. We are
in fact going to show more, namely that the Lagrangian manifolds
(\textquotedblleft invariant tori\textquotedblright ) $\mathbb{T}$
associated to $H$ automatically satisfy the \textit{EBK} quantum condition%
\begin{equation}
\frac{1}{h}\doint\nolimits_{\gamma }pdq-\frac{1}{4}\mu (\gamma )\text{ \ 
\textit{is an integer}}\geq 0  \label{EBK}
\end{equation}%
for all \textit{loops}\footnote{%
It is sometimes mistakingly believed that the $\gamma $ have to be periodic
orbits of the Hamiltonian itself.} $\gamma $ drawn on $\mathbb{T}$ ($\mu
(\gamma )$ is the Maslov index of $\gamma $; see e.g \cite{ICP,Gutzwiller}).
The semiclassical approximation to the quantum energy is then obtained by
calculating the energy along the classical trajectories of $H$ satisfying (%
\ref{EBK}). The first step of the proof consists in introducing action-angle
variables $(\phi ,I)=(\phi _{1},...,\phi _{N};I_{1},...,I_{N})$. Hamilton's
function becomes $K(I)=H(p,q)$ and the corresponding equations of motion are%
\begin{equation*}
\dot{\phi}_{j}=\frac{\partial K(I)}{\partial I}\equiv \omega _{j}(I)\text{ \
\ , \ \ }\dot{I}_{j}=0\text{ \ \ , \ \ }j=1,...,N\text{.}
\end{equation*}%
In these coordinates $\mathbb{T}$ becomes a new Lagrangian manifold $\mathbb{%
T}_{\phi ,I}$ defined by the condition $I=I(0)=$\textit{constant}, and can
be parametrized by the angles $\phi _{j}$. The next step is to define a new
Hamiltonian function by 
\begin{equation*}
K_{0}(I)=\omega _{1}(0)I_{1}+\cdot \cdot \cdot +\omega _{N}(0)I_{N}
\end{equation*}%
where we have set $\omega _{j}(0)=\omega _{j}(I(0))$ for $1\leq j\leq N$.
The trajectory for $K_{0}$ passing through $(\varphi (0),I(0))$ at time $t=0$
will lie on $\mathbb{T}_{\phi ,I}$. We now make a new canonical change of
variables $(\phi ,I)\longmapsto (Q,P)$ where the $Q_{j},P_{j}$ are obtained
from the $\phi _{j},I_{j}$ as the modified polar coordinates%
\begin{equation*}
Q_{j}=\sqrt{2I_{j}}\cos \phi \text{ \ \ , \ \ }P_{j}=\sqrt{2I_{j}}\sin \phi 
\text{.}
\end{equation*}%
This transformation brings the Hamiltonian $K_{0}$ into the form%
\begin{equation}
H_{0}(Q,P)=\frac{\omega _{1}(0)}{2}(P_{1}^{2}+Q_{1}^{2})+\cdot \cdot \cdot +%
\frac{\omega _{N}(0)}{2}(P_{N}^{2}+Q_{N}^{2})  \label{ho}
\end{equation}%
and $\mathbb{T}_{\phi ,I}$ becomes in these coordinates the torus $\mathbb{T}%
_{Q,P}=C_{1}\times \cdot \cdot \cdot \times C_{N}$, where the $C_{j}$ are
the circles 
\begin{equation*}
C_{j}:P_{j}^{2}+Q_{j}^{2}=\omega _{j}(0)I_{j}(0)
\end{equation*}%
lying in the $Q_{j},P_{j}$ plane. Let now $\gamma $ be an orbit for the
Hamiltonian $H_{0}$, starting from some point of $\mathbb{T}_{Q,P}$. That
orbit will not only wind around $\mathbb{T}_{Q,P}$ but also around each
cylinder $Z_{j}(\sqrt{\omega _{j}(0)I_{j}(0)})$; in view of the quantum blob
Ansatz that cylinder must be a quantum blob if the motion is semiclassically
admissible, and hence%
\begin{equation}
\pi \omega _{j}(0)I_{j}(0)=(n_{j}+\tfrac{1}{2})h  \label{key}
\end{equation}%
for some integer $n_{j}\geq 0$. Choose a topological basis $\epsilon
_{1}^{\prime },...,\epsilon _{N}^{\prime }$ of $\mathbb{T}_{Q,P}$ consisting
of the circles $C_{j}$ parametrized as $Q_{jk}(t)=P_{jk}(t)=0$ if $k\neq j$
and%
\begin{equation*}
Q_{jj}(t)=\omega _{j}(0)I_{j}(0)\cos t\text{ \ , \ }P_{jj}(t)=\omega
_{j}(0)I_{j}(0)\sin t\text{ \ }(\ 0\leq t\leq 2\pi )
\end{equation*}%
for $j=1,...,N$; we have 
\begin{equation}
\doint\nolimits_{\epsilon _{j}^{\prime }}PdQ=\doint\nolimits_{\epsilon
_{j}^{\prime }}P_{j}dQ_{j}=(n_{j}+\tfrac{1}{2})h\text{.}  \label{BSE}
\end{equation}%
Returning to the original coordinates $q,p$ and denoting by $\epsilon
_{1},...,\epsilon _{N}$ the topological basis $\epsilon _{1}^{\prime
},...,\epsilon _{N}^{\prime }$ expressed in these coordinates, we finally get%
\begin{equation*}
\doint\nolimits_{\epsilon _{j}}pdq=\doint\nolimits_{\epsilon _{j}^{\prime
}}PdQ=(n_{j}+\tfrac{1}{2})h
\end{equation*}%
since action integrals around loops are invariant under canonical
transformations. Hence 
\begin{equation}
\doint\nolimits_{\gamma }pdq=\sum_{j=1}^{N}\nu _{j}\doint\nolimits_{\epsilon
_{j}}pdq=h\sum_{j=1}^{N}\nu _{j}n_{j}+\frac{1}{2}h\sum_{j=1}^{N}\nu _{j}%
\text{;}  \label{sum}
\end{equation}%
since the Maslov index is, by definition, 
\begin{equation*}
\mu (\gamma )=2\sum_{j=1}^{N}\nu _{j}
\end{equation*}%
formula (\ref{sum}) implies that 
\begin{equation*}
\frac{1}{h}\doint\nolimits_{\gamma }pdq=\frac{1}{4}\mu (\gamma )+\text{%
\textit{integer}}
\end{equation*}%
which is precisely the announced \textit{EBK} condition (\ref{EBK}).

\section{Concluding Remarks}

All the existing proofs of Gromov's theorem are difficult and make use of
sophisticated mathematical techniques. It is probably the reason why it has
taken two hundred years after Lagrange\footnote{%
We recall that Lagrange proposed the letter $H$ in his \textit{M\'{e}%
chanique Analytique }to honor Huygens --not Hamilton who still was in his
early childhood at that time!} wrote down \textquotedblleft Hamilton's
equations\textquotedblright\ to discover the principle of the symplectic
camel! We refer to \cite{EH,Gromov,HZ,Viterbo} for detailed proofs; a
heuristic justification is given in \cite{ICP}.

In the present state of the art symplectic areas are very difficult to
calculate explicitly, outside a few particular cases (one of which being the
situation of the double inclusion (\ref{inclus})). The notions introduced in
this Letter might therefore be for the moment of a more theoretical than
practical value. We hope that they might provide some insights in related
fields of current research such as Bose-Einstein condensation (in which we
obtain a single quantum state: for a \textit{BEC} the notion of
\textquotedblleft number of particles\textquotedblright\ does not make
sense, so it might be viewed as a gigantic quantum blob immersed in some
(indefinite dimensional) phase space), or in a better understanding of the
Casimir effect (which is related to renormalization questions for the ground
energy levels). A natural, less grandiose, application would of course be to
find out whether the quantum blob Ansatz would be of some utility in the
study of semiclassical quantization of non-integrable Hamiltonian systems.
This is plausible, because symplectic area seems to be related to the notion
of adiabatic invariance, and one could then envisage applying it to the
method of adiabatic switching (see \cite{SBR}; also \cite{BOG} in the
context of ergodicity).

\end{document}